\documentstyle[12pt]{article}

\begin{document}
\thispagestyle{empty} \vspace{-2.0cm}
\bigskip
\begin{center}
{\Large \bf Translational groups as   generators of gauge transformations}
\vskip 0.4cm
{\bf Tomy Scaria}\footnote{Present Address: Theory Division, Saha Institute of Nuclear Physics, 1/AF Bidhan Nagar, Calcutta 70064. $~~~~~~~~~~~~~~~~~~~~~~~~~~~~~~$e-mail: {\it toms@theory.saha.ernet.in}}
\vskip 0.4 true cm
S. N. Bose National Centre for Basic Sciences \\
JD Block, Sector III, Salt Lake City, Calcutta -700 098, India.\\
\end{center}
\vskip 0.3true cm
\centerline{\large \bf Abstract}
We examine the gauge generating nature of the translational subgroup of
Wigner's little group for the case of massless
 tensor
gauge theories and show that the gauge transformations generated by the translational group is only
a subset of the complete set of gauge transformations.
We  also show that, just like the case of topologically massive gauge theories,
 translational groups act as generators of gauge transformations in gauge theories obtained by extending
 massive gauge noninvariant theories by a
 St\"uckelberg mechanism. The  representations of the translational groups
that generate gauge transformations in such
 St\"uckelberg extended theories can
be obtained by the method of dimensional descent. We illustrate these with the examples of
St\"uckelberg extended first class versions of Proca, Einstein-Pauli-Fierz and
massive Kalb-Ramond theories  in 3+1 dimensions. A detailed analysis of the
partial gauge generation in massive and massless 2nd rank symmetric gauge
theories is provided. The gauge transformations generated by translational
group in 2-form gauge theories are shown to explicitly manifest the
reducibility of gauge transformations in these theories.
\vskip 0.5cm
\begin{flushleft}
{\large PACS No.(s)} 11.15.-q,  11.10.Kk,  04.20.Cv \\
{\large Keywords:} translational group, gauge transformation
\end{flushleft}

\section{Introduction}

Wigner's little group is quite familiar to physicists mainly
because of  its role in the classification of elementary particles.
Wigner introduced the concept
of little group in a seminal paper \cite{w} published in 1939 and showed
how particles can be classified on the basis of their spin/ helicity quantum
numbers using the little group.   The little group   also   relates the internal symmetries of massive
 and massless particles \cite{kim}.   A comparatively lesser
known facet of the little group is  its role as a generator of gauge
transformations in various Abelian gauge theories. This aspect of little group was  first noticed in
the contexts  of free Maxwell theory  \cite{we,hk,hk1} and linearized Einstein gravity \cite{ng}.
Recently, it was shown that the little group for massless particles  acts as generator of gauge
transformations  in the case of other gauge theories as well \cite{b}. For example,
the  defining representation of this   little group  is shown to
generate gauge transformations also in  the 3+1 dimensional Kalb-Ramond(KR) theory which is a massless
2-form gauge theory  \cite{bc1,rpm}.
 To be precise, it is the translational subgroup $T(2)$\footnote{Group of translations in $n$ dimensional
space is denoted by $T(n)$.} of the Wigner's
little group for massless particles   that generate gauge transformations in these theories.
On the other hand, in 3+1 dimensional $B \wedge F$ theory which is a topologically
massive gauge theory, one need to go beyond $T(2)$ and it is a particular representation of the
translational group $T(3)$ that generate the gauge transformations in this
theory. However, as shown in \cite{bc2}, one can easily see that $T(3)$ is a
subgroup of Wigner's little group    for a massless particle in 4+1 dimensions
which generates gauge transformations
in massless theories living in this  higher dimensional space-time. It is further
shown in \cite{bc2} that one can systematically derive the representation of
$T(3)$ that acts as gauge generator in   3+1 dimensional $B \wedge F$ theory
from the gauge transformation properties of free Maxwell and KR
theories in 4+1  dimensions using a method called `dimensional descent'.
Similarly, dimensional descent can also be employed to obtain the
gauge generating representations of $T(1)$ for topologically massive Maxwell-Chern-Simons
and linearized Einstein-Chern-Simons gauge theories \cite{djt,bcs1,bcs} in 2+1 dimensions by
starting respectively from Maxwell and linearized gravity theories in 3+1
dimensions \cite{bc2,sc}.

Against this background, the purposes of  the present study are the following.
Firstly, we make a closer analysis of the gauge
generation by Wigner's little group in massless tensor gauge theories, namely
linearized gravity and  Kalb-Ramond(KR) theories, and  unravel certain
subtle points which went unnoticed in earlier studies.
We show that  the translational group  $T(2)$  generates only a subset of the full
range of gauge transformations in these   theories.
Furthermore, in the case of KR and $B \wedge F$
theories,  the generators of gauge transformations
are not  all  independent and  such theories are known as  `reducible gauge systems' \cite{gomis,bn}.
Our analysis shows that gauge generation
by the translational group $T(2)$ in a reducible gauge theory  manifestly exhibits
the reducibility  the  gauge transformations (section 3).

Secondly, one  should note that, apart from the usual massless gauge theories
and topologically massive gauge
theories,  there exists gauge theories that can be obtained by   converting second class
constrained systems (in the  language of Dirac's theory of constraint dynamics) to first
class (gauge) systems using the generalized   canonical
prescription of Batalin, Fradkin and Tyutin \cite{bft}. By such a prescription one can obtain
from the massive gauge noninvariant  theories, their St\"uckelberg extended versions  which are massive
as well as gauge invariant \cite{bb}. Now one may wonder if  translational groups
act as gauge generators in such massive gauge theories as well. If so,
what would be the representation of these groups that generate such gauge
transformations?   In the present study, we delve in to
these questions and  show that the same representation of the translational
group $T(3)$
that generates gauge transformation in  the topologically massive $B\wedge F$ theory
 also generates gauge
transformation in the St\"uckelberg extended     versions of
 Proca,  Einstein-Pauli-Fierz (EPF) and massive KR theories (section 4).

Finally,
it  will  be shown that, just as in topologically massive gauge theories, dimensional descent  can also be used
to obtain the polarization vectors (or tensors) and momentum vectors of the St\"uckelberg
models and the gauge generating  representation of translational
group in such models, by starting
from appropriate theories in  one higher space-time dimension (section 5).

{\bf Notation:}  We use  $\mu, \nu $ etc.
for denoting  indices in 3+1 dimensional space-time. The letters
$i,j$ etc.are used for 4+1 dimensional space-time except in section 4
where they represent spatial components of 3+1 dimensional vectors/tensors. Metric used is mostly negative.
We denote polarization vectors by $\varepsilon_{\mu}$ and polarization tensors
of 2-form theories are denoted by $\varepsilon_{\mu \nu}$ while those of  symmetric 2nd
rank tensor fields   by $\chi_{\mu\nu}$.

\section{Wigner's little group as a generator of gauge transformations in
various theories}

Historically, the gauge generating property of the little group was first
 studied in  the context of free Maxwell theory \cite{we,hk,hk1} where
it was shown that the  action of  little group on the polarization vector
of Maxwell photons amounts to gauge transformation in momentum space.
As is well known, free Maxwell theory is described by the
Lagrangian
\begin{equation}
{\cal{L}} = -\frac{1}{4}F_{\mu \nu }F^{\mu \nu }, ~~~~~F_{\mu \nu } =
\partial_\mu A_\nu - \partial_\nu A_\mu
\label{7}
\end{equation}
which is invariant under the gauge transformation $
A_{\mu}(x) \rightarrow  A_{\mu}(x) + \partial_{\mu}\tilde{f}(x) $
where $\tilde{f}(x)$ is an arbitrary scalar function.
The Lagrangian (\ref{7}) leads to  the
equation of motion; $
\partial_\mu F^{\mu \nu} = 0. $
Denoting the polarization vector of a photon by $\varepsilon^{\mu}(k)$,  a solution of this equation
can
be written as
\begin{equation}
A^{\mu}(x) = \varepsilon^{\mu}(k) e^{ik \cdot x}
\label{9}
\end{equation}
where only a single mode is considered and  the positive frequency part is suppressed  for simplicity.
In terms of the
polarization vector $\varepsilon^{\mu}(k)$, the gauge transformation  and the equation of motion for Maxwell theory
are expressed respectively as   follows;
\begin{equation}
\varepsilon_{\mu}(k) \rightarrow \varepsilon^{\prime}_{\mu}(k) =  \varepsilon_{\mu}(k) + if(k)k_{\mu}
\label{10}
\end{equation}
 \begin{equation}
k^2 \varepsilon^{\mu} - k^{\mu} k_{\nu} \varepsilon^{\nu} = 0
\label{11}
\end{equation}
where $\tilde{f}(x)$ has been written as $\tilde{f}(x) = f(k)e^{ik \cdot x}$.
The massive excitations corresponding to $k^2 \neq 0$ leads to a solution
$\varepsilon^{\mu} \propto k^{\mu}$ which can therefore be gauged away by a
suitable choice of $f(k)$ in (\ref{10}). For
massless excitations  ($k^2 = 0$), the Lorentz condition $
k_\mu \varepsilon^{\mu} =0$
 follows immediately
 from (\ref{11}).  For a photon of energy $\omega$ propagating in the $z$-direction, the 4-momentum can be written as $k^{\mu} = (\omega, 0, 0, \omega)^T$.
It then follows from ({\ref{11}) that the corresponding polarization tensor   $\varepsilon^{\mu}(k)$ takes
the form $(\varepsilon^{0}, \varepsilon^1, \varepsilon^2, \varepsilon^{0})$ which can be reduced to the
maximally reduced form\footnote{This procedure of obtaining the
maximally reduced polarization vectors/tensors of
various theories by choosing a plane wave solution for the respective equation motion will henceforth be referred
 to as the ``plane wave method".}
\begin{equation}
\varepsilon^{\mu}(k) =
(0, \varepsilon^1, \varepsilon^2, 0)^T
\label{13}
\end{equation}
by a   gauge transformation (\ref{10}) with $f(k) = \frac{i\varepsilon^0}{\omega}$. Note that the maximally reduced
form (\ref{13})
of $\varepsilon^{\mu}$ displays just the two physical degrees of freedom
$\varepsilon^1$ and $\varepsilon^2$.

We now   digress briefly to recapitulate the essential aspects of
Wigner's little   group.  The Wigner's little group ${\cal W}$ is defined as
the subgroup of homogeneous Lorentz group $L$ that preserves the
 energy-momentum vector of a particle:
\begin{equation}
{{\cal W}^\mu}_\nu k^\nu = k^\mu
\label{1}
\end{equation}
where $k^\mu$ is an arbitrary, but  fixed, vector on the mass shell $ {\cal M}_{m^2} = \{ k^\mu | k^2 = m^2\} $. ${\cal M}_{m^2}$ is acted on transitively
by the Lorentz group $L$. Little group $W(k)$ is the stability subgroup of $L$ so that $ {\cal M}_{m^2}$ can be identified  as a homogeneous
coset space $L/{\cal W} $.
It is obvious that, in 3+1 dimensions,  the little group for a massive particle
is the  rotational group $SO(3)$. On the other hand, for a
massless  particle, the little group is isomorphic to the Euclidean group $E(2)$ which is a
semi-direct product of $SO(2)$ and $T(2)$ - the group of translations in the
2-dimensional plane \cite{hk1}. The  explicit representation of Wigner's
little group ${\cal W}_4$ that preserves the 4-momentum $k^\mu = (\omega, 0, 0, \omega)^T$ of a photon
of energy $ \omega $ moving in the $z$-direction is given by \cite{we}
\begin{equation}
{\cal W}_4(p, q;  \phi) =
\left( \begin{array}{cccc}
1+ \frac{p^2 + q^2 }{2} & p\cos \phi - q \sin \phi & p\sin \phi + q \cos \phi  & -\frac{p^2 + q^2 }{2} \\
p & \cos \phi & \sin \phi  & -p \\
q & -\sin \phi & \cos \phi  & -q \\
\frac{p^2 + q^2 }{2} &  p\cos \phi -q \sin \phi & p\sin \phi + q \cos \phi  & 1
-\frac{p^2 + q^2 }{2}
\end{array}\right).
\label{2}
\end{equation}
Here $p$ and $q$ are   real parameters.
This little group  can be written as $W_4(p, q;  \phi) = W(p, q) R(\phi) $ where
\begin{equation}
W(p, q) \equiv {\cal W}_4(p, q;  0) = \left( \begin{array}{cccc}
1+ \frac{p^2 + q^2 }{2} & p & q  & -\frac{p^2 + q^2 }{2} \\
p & 1 & 0 & -p \\
q & 0 & 1 & -q \\
\frac{p^2 + q^2 }{2} &  p & q  & 1 -\frac{p^2 + q^2 }{2}
\end{array}\right)
\label{4}
\end{equation}
is  a particular representation of the  translational subgroup  $T(2)$ of
the little group and $ R(\phi)$ represents a $SO(2)$ rotation about the $z$-axis.
Note that the representation $W(p, q)$ satisfies the  relation $ W(p, q)W(\bar{p},\bar{q}) = W(p+\bar{p}, q+\bar{q})$.

Under the action  of the translational group $T(2)$ in (\ref{4}),  the 
maximally reduced polarization
vector (\ref{13}) of Maxwell theory transforms  as follows:
\begin{equation}
\varepsilon^{\mu} \rightarrow \varepsilon^{\prime \mu} = {W^{\mu}}_{\nu}(p, q) \varepsilon^{\nu} = \varepsilon^{\mu} +
 \left( \frac{p\varepsilon^1 + q\varepsilon^2}{\omega}\right)k^{\mu}~.
\label{14}
\end{equation}
Clearly, this can be identified as a gauge transformation of the form of 
(\ref{10}) by choosing $f(k) = \frac{p\varepsilon^1 + q\varepsilon^2}{i\omega}$ 
 thus displaying the gauge generating property
of Wigner's little group for massless particles in free Maxwell theory. Conversely, any general gauge transformation 
(in momentum space) in Maxwell theory can be viewed as resulting from the action of translational
group $W(p, q)$ on the polarization vector of the theory.

The same translational group $T(2)$ in the representation (\ref{4}) generates gauge
transformations in 3+1 dimensional linearized gravity \cite{sc} and Kalb-Ramond
theory \cite{bc1}. However, as we will show in the next section, the transformations generated by
 $T(2)$ is only a subset of the whole spectrum  of gauge
transformations in these massless tensor gauge
theories.
On the other hand, in the case of the topologically massive $B\wedge F$ gauge theory \cite{al},
 the translational
 group (\ref{4}) fails to act as gauge generator. The generator of gauge
transformation in $B\wedge F$ theory is shown in \cite{bc1}
to be the translational group $T(3)$ in the representation
\begin{equation}
D(p, q, r) = \left( \begin{array}{cccc}
1 & p & q & r \\
0 & 1 & 0 & 0 \\
0 & 0 & 1 & 0 \\
0 & 0 & 0 & 1
\end{array} \right)
\label{53-1}
\end{equation}
(where $p, q, r$ are real parameters). 
Another representation  of $T(3)$, inherited from the defining representation of
 Wigner's little group for massless particles in 4+1 dimensions,
generates gauge transformation in 4+1 dimensional Maxwell theory.
The close relationship between these different representations of $T(3)$ is
analyzed in detail in \cite{bc2}   using the method called dimensional descent which will be discussed in section 5
in the present context.

\section{Partial gauge generation by $T(2)$ in massless tensor gauge theories}

It was argued respectively in \cite{bc1} and \cite{ng,sc} that the translational
group $T(2)$  in the representation (\ref{4}) generates gauge transformations
in 3+1 dimensional massless KR theory and linearized gravity.  As pointed out in \cite{ng}
the gauge generation by $T(2)$ in linearized gravity is subject to
certain restrictions. Here  we make a closer examination of this partial gauge generation by $T(2)$
in linearized gravity  revealing some aspects which went unnoticed before. We
also show that $T(2)$ generates only a restricted set of gauge transformations
in massless KR theory and that the reducible nature of its gauge transformations are
reflected in this partial gauge generation, points that were missed  in earlier studies.

 We first consider linearized gravity\footnote{In linearized gravity, the metric $g_{\mu\nu}$ is
assumed to be close to the  flat background part $\eta_{\mu\nu}$ and one 
writes $g_{\mu\nu} = \eta_{\mu\nu} + h_{\mu\nu}$ with deviation $ |h_{\mu\nu}|<< 1$. 
Raising and lowering of indices are done by $\eta^{\mu\nu}$ and $\eta_{\mu\nu}$ respectively.} which is governed by the 
Lagrangian
\begin{equation}
{\cal L}_L^E = \frac{1}{2}h_{\mu\nu} \left[ R^{\mu\nu}_L - \frac{1}{2} \eta^{\mu\nu} R_L\right];~~~ R^{\mu\nu}_L = \frac{1}{2}(- \Box h^{\mu\nu} + \partial^\mu \partial_\alpha  h^{\alpha \nu} + \partial^\nu\partial_\alpha  h^{\alpha \mu} - \partial^\mu \partial^\nu h)
\label{44}
\end{equation}
where $ R^{\mu\nu}_L$ is the linearized Ricci tensor while
 $h = h^\alpha_\alpha$ and $R_L = {R^\alpha_L}_\alpha$.
Linearized gravity is
 invariant under the gauge transformation
\begin{equation}
h^{\mu\nu}(x) \rightarrow h'^{\mu\nu}(x) = h^{\mu\nu}(x) + \partial^\mu \tilde{\zeta}^\nu (x) 
+ \partial^\nu  \tilde{\zeta}^\mu (x).
\label{47}
\end{equation} 
Adopting the ansatz (analogous to (\ref{9}))
$h^{\mu\nu} = \chi^{\mu\nu} (k) e^{ik.x}  $,
where $\chi^{\mu\nu}$ is the symmetric  polarization tensor,
the gauge transformation (\ref{47}) can be written in the momentum space as
\begin{equation}
 \chi^{\mu\nu} (k)  \rightarrow {\chi}'^{\mu\nu} (k) = \chi^{\mu\nu} (k) + k^\mu \zeta^\nu (k) +  k^\nu \zeta^\mu (k)
\label{50}
\end{equation}
    with
$\tilde{\zeta}^\mu (x) = \zeta^\mu (k) e^{ik.x}$.
Now, following the plane wave method as described in \cite{sc}, one can obtain  the
maximally reduced form of polarization tensor corresponding to a particle with the
4-momentum $k^\mu = (\omega , 0,0, \omega )^T$ and is given by
\begin{equation}
 \{\chi^{\mu\nu}\} = \left(
\begin{array}{cccc}
0 & 0 & 0 & 0\\
0 & a & b & 0 \\
0 & b & -a & 0 \\
0 & 0 & 0 & 0
\end{array}
\right).
\label{53}
\end{equation}
(For another derivation see \cite{weinberg1}.)
Here $a$ and $b$ are are free parameters  representing the two physical
degrees of freedom in 3+1-dimensional linearized gravity\footnote{Linearized gravity in $d$ dimensions has $\frac{d(d-3)}{2}$ degrees of freedom \cite{weinberg1}.}.
Notice that the maximally reduced form (\ref{53}) of the polarization tensor satisfies the
momentum space harmonic gauge condition,
$k_\mu \chi^{\mu}_\nu = \frac{1}{2}k_\nu \chi^{\mu}_\mu $    \cite{weinberg1}.

It is now easy to show that the action of the translational group
$W(p, q) $ (\ref{4}) on the polarization tensor (\ref{53})  is equivalent
to a gauge transformation.
$$\{ \chi^{\mu \nu}\} \rightarrow \{\chi'^{\mu\nu}\}= W(p, q) \{\chi^{\mu\nu}\} W^T(p, q)$$
\begin{equation}
 = \{\chi^{\mu\nu}\} +\left(
\begin{array}{cccc}
((p^2 - q^2)a + 2pqb)  & (pa + qb) & (pb -qa) & ((p^2 - q^2)a + 2pqb) \\
(pa + qb) &0 & 0 & (pa + qb) \\
(pb -qa)  & 0 & 0 & (pb -qa) \\
((p^2 - q^2)a + 2pqb) &( pa + qb) &( pb -qa) & ((p^2 - q^2)a + 2pqb)
\end{array}\right).
\label{544}
\end{equation}
The above transformation can be cast in the form of a gauge transformation
(\ref{50}) with the
following choice for  the arbitrary functions $\zeta^\mu(k)$ \cite{sc}:
\begin{equation}
\zeta^1 = \frac{pa + qb}{\omega},~~~~\zeta^2 = \frac{pb -qa}{\omega},~~~~
\zeta^0 = \zeta^3 = \frac{(p^2 - q^2)a + 2pqb}{2\omega}.
\label{555}
\end{equation}
However, since $k^\mu = (\omega , 0,0, \omega )^T$,  a general gauge
transformation  for $ \{\chi^{\mu\nu}\}$ (\ref{53}) has the form
\begin{equation}
\{\chi^{\mu\nu} \} \rightarrow  \{\chi^{\mu\nu} \} + \{ k^\mu \zeta^\nu \} +
\{k^\nu \zeta^\mu\}
= \{\chi^{\mu\nu}\} +\omega\left(
\begin{array}{cccc}
2 \zeta^0   & \zeta^1 & \zeta^2  & (\zeta^0 + \zeta^3) \\
\zeta^1 &0 & 0 & \zeta^1 \\
\zeta^2  & 0 & 0 & \zeta^2 \\
(\zeta^0 + \zeta^3) &\zeta^1 &\zeta^2 & 2 \zeta^3
\end{array}\right).
\label{577}
\end{equation}
Upon comparing the above form of general gauge transformation with the one
generated by $W(p,q)$ given in (\ref{544}), it becomes clear that the latter is
only a special case of the former as the relations in
(\ref{555})  restricts the number of independent components of the arbitrary 
vector  $ \zeta^\mu$. Therefore, the translational subgroup $T(2)$ of
Wigner's little  group for massless particles generates only a subset of the 
full set of gauge transformations in linearized gravity.  In this connection one must notice that
 the gauge freedom in linearized gravity is represented by
the arbitrary vector variable $ \zeta^\mu$ having four components, while
the translational group   $T(2)$ has only two parameters. Naturally,
in the gauge generation by $W(p,q)$ in linearized gravity, only two of the four
components of $ \zeta^\mu$  remain independent (as is evident from 
(\ref{577}))  when expressed in terms of the two parameters $(p,q)$
and therefore the gauge generation is only partial.
It was noted in \cite{ng} that the gauge generation by the little group in 
linearized gravity is subject to the `Lorentz condition' $k_\mu \zeta^\mu (k) = 0$.
This also can be seen from the third relation $\zeta^0 = \zeta^3$ in
(\ref{555}) since $k^\mu = (\omega , 0,0, \omega )^T$.
Thus, our present analysis  have unraveled all the constraints
behind the partial gauge generation by Wigner's little group in linearized 
gravity. In contrast, the gauge freedom in free Maxwell theory is represented
by a single arbitrary scalar variable $f(k)$ (\ref{10}) which can be expressed 
(without any restrictions) in
terms of the two parameters of  $W(p,q)$ in the gauge generation by little
group as is evident from (\ref{14}). 
Hence translational subgroup of Wigner's little group  generates the full set of gauge 
transformations in Maxwell theory.

We now consider the gauge transformations generated by the translational group $W(p,q)$ 
in massless KR theory which has a second rank antisymmetric tensor
as its basic field. The KR theory is described by the Lagrangian,
\begin{equation}
{\cal L} = \frac{1}{12}H_{\mu\nu\lambda}H^{\mu\nu\lambda}; ~~~~H_{\mu\nu\lambda} = 
\partial_\mu B_{\nu\lambda} + \partial_\nu B_{\lambda\mu} + \partial_\lambda B_{\mu\nu}
\label{199}
\end{equation}
where $B_{\mu\nu}$ is the 2nd rank antisymmetric gauge field ($B_{\mu\nu} = - B_{\nu\mu}$).
The KR theory is invariant under the gauge transformation
\begin{equation}
B_{\mu\nu}(x) \rightarrow   B'_{\mu\nu}(x) = B_{\mu\nu}(x) + \partial_\mu F_\nu (x) - \partial_\nu F_\mu (x)
\label{1993}
\end{equation}
where $F_\mu (x)$ are arbitrary functions.
One can see that under the transformation
\begin{equation}
F_\mu (x)
 \rightarrow F'_\mu (x) = F_\mu (x) + \partial_\mu \beta (x)
\label{1993+1}
\end{equation}
 (where $\beta (x)$ is
an arbitrary scalar function) the gauge transformation (\ref{1993}) remains invariant.
In particular, if $F_\mu = \partial_\mu \Lambda$  the gauge transformation vanish trivially.
This is known as  the `gauge invariance of gauge transformations' and is  a typical
property of reducible gauge theories\footnote{Notice a crucial difference in the
 case of
linearized gravity which has the  symmetric tensor $h_{\mu\nu}$ as its underlying gauge field.
Under a transformation of the type (\ref{1993+1}),
 the gauge transformation (\ref{47}) changes.
This shows that, unlike KR theory, there is no `gauge invariance of gauge transformation' in
linearized gravity which is not a reducible gauge system.} where the generators of gauge transformation
are not all independent \cite{gomis}.    Hence
 there exists some superfluity in the
 gauge transformation (\ref{1993}).
The maximally reduced form of the antisymmetric polarization tensor $\varepsilon^{\mu \nu}$
associated with the 2-form potential $B^{\mu \nu} (= \varepsilon^{\mu \nu} e^{ik.x}$) of
KR theory is obtained in \cite{bc1} using plane wave method as
\begin{equation}
\{\varepsilon^{\mu\nu}\} = \varepsilon^{12}
  \left( \begin{array}{cccc}
0 & 0 & 0 & 0 \\
0 & 0 & 1 & 0 \\
0 & -1 & 0 & 0 \\
0 & 0 & 0 & 0
 \end{array} \right).
\label{19}
\end{equation}
Notice that, similar to  Maxwell and linearized gravity theories,  this form of the polarization tensor satisfies a `Lorentz condition' $
k_\mu \varepsilon^{\mu \nu} = 0 $
corresponding to $\partial_\mu B^{\mu \nu} = 0$. 
In the case of KR theory, the counterpart of the momentum space gauge 
transformation (\ref{50}) is given by 
\begin{equation}
\varepsilon^{\mu \nu}(k) \rightarrow \varepsilon'^{\mu \nu}(k) = \varepsilon^{\mu \nu}(k) + i(k^{\mu}f^{\nu}(k) - k^{\nu}f^{\mu}(k))
\label{18+1}
\end{equation}
where $f_{\mu}(k)$ are arbitrary and independent functions of $k$ (with $F_\mu (x) = f_\mu (k) e^{ik.x}$).
The transformation of $ \{\varepsilon_{\mu\nu}\}$ 
under the translational group $W(p, q)$ (\ref{4}), can be written as
\begin{equation}
\{\varepsilon ^{\mu \nu} \}\rightarrow \{\varepsilon'^{\mu \nu}\}  = W(p,q) \{\varepsilon^{\mu \nu}\}  W^T (p,q) =
 \{\varepsilon^{\mu \nu} \}+ \varepsilon^{12}\left( \begin{array}{cccc}
0 & -q & p & 0 \\
q & 0& 0 & q \\
-p & 0 & 0 &  -p \\
0 & -q & p & 0
\end{array} \right)
\label{20}
\end{equation}
This can be cast in the
form of  (\ref{18+1}) with
\begin{equation}
f^1 = \frac{-q\varepsilon^{12}}{i\omega},~~~
f^2 =  \frac{p\varepsilon^{12}}{i\omega}, ~~~
f^3 = f^0.
\label{2001}
\end{equation}
 As in the case of linearized gravity, on account of the 
requirement $f^3 = f^0$, the gauge transformations generated by the 
translational group fails to include the entire set of gauge transformations
in KR theory too. Analogous to (\ref{577}), the general form of gauge 
transformation (\ref{18+1}) in the matrix form is
\begin{equation}
\{\varepsilon^{\mu \nu} \} \rightarrow \{\varepsilon'^{\mu \nu} \} = \{\varepsilon^{\mu \nu}\}  + \omega \left( \begin{array}{cccc}
0 & f^1 & f^2 & f^0 - f^3 \\
-f^1 & 0& 0 &  -f^1\\
-f^2 & 0 & 0 &  - f^2 \\
 f^3  -  f^0 & f^1 &  f^2 & 0
\end{array} \right)
\label{200}
\end{equation}
which makes it quite explicit that the transformation (\ref{20})
 does not exhaust (\ref{200}), but is only a special case  (where $f^0 = f^3$) of it.
The transformation (\ref{20}) is an attempt to generate the gauge equivalence class
of the maximally reduced polarization tensor (\ref{19})
of KR theory using only the two parameters of the translational group $W(p,q)$
while the full gauge freedom of the theory is represented by the arbitrary
4-vector variable $f^\mu(k)$. Hence analogous to the case of linearized gravity, the gauge generation by $W(p,q)$
in massless KR theory is only partial.
Moreover,  here also the arbitrary  function $f^\mu(k)$
satisfy the `Lorentz condition' $k_\mu f^\mu(k) = 0$ since $f^0 = f^3$ (\ref{2001}).

It is important to notice that in (\ref{2001}), while the components $f^1$ and $f^2$ of
$f^\mu$ are expressed in terms of the parameters $p, q$ of the
translational group $W(p,q)$,  the other two components ($f^0, f^3$)
are independent of the parameters (and of the maximally reduced polarization
 tensor) and are left completely undetermined subject only
to the  constraint $f^0 = f^3$.
Thus, in the gauge generation by $W(p,q)$ in KR theory, corresponding to any given pair
($f^1, f^2$) there exists a continuum of allowed choices for $f^0 (= f^3$)
representative of the invariance of gauge transformations (\ref{1993})
under (\ref{1993+1}). Therefore, the partial gauge generation by $W(p,q)$
in massless KR theory clearly exhibits the reducibility of its gauge transformations.
 This may be compared to
the gauge generation (\ref{544}) in linearized gravity by $W(p,q)$ where all the components of
the arbitrary vector variable $\zeta^\mu$  are expressed in terms of the
parameters $(p,q)$ (see (\ref{555})) hence indicating the absence of any reducibility
in the gauge transformation of the theory.

Notice that the transformation (\ref{1993+1}) is of same form as the
gauge transformation  in Maxwell theory where $W(p,q)$ acts as gauge  generator.
Hence,  the `gauge
transformation (\ref{1993+1}) of gauge transformations' in KR theory   may be considered
as being generated by  the translational group $W(p,q)$. Thus, in KR theory which is a 2-form gauge theory,
  two independent elements of the translational group $W(p,q)$ are involved
in generating gauge transformations, one for the underlying 2-form field
$B_{\mu\nu}$ and the other for the field $F_\mu$ which correspond to the gauge 
freedom of the theory.
In the gauge generation for massless theories by the translational group
$W(p,q)$, we therefore perceive an appealing
hierarchical structure starting from the  Maxwell (1-form)
and KR (2-form) theories; namely  in a $n$-form theory, $n$  elements of the
translational group $W(p,q)$ being involved in gauge
generation. It is expected that this hierarchical structure continues for higher form gauge theories as well.

\section{Massive gauge theories}

In this section we study the relationship between the translational groups and gauge transformation
in gauge theories which are obtained from massive theories through  St\"uckelberg mechanism.
Only 3+1 dimensional
theories are considered in this section.
\subsection{Massive vector gauge theory}
One can render the 3+1-dimensional Proca theory (which does not possess any  gauge symmetry) gauge invariant
by  St\"uckelberg mechanism with the introduction of a new scalar field  $\alpha
(x)$
as follows;
\begin{equation}
{\cal{L}} = -\frac{1}{4}F_{\mu \nu }F^{\mu \nu } + \frac{m^2}{2} (A_\mu + \partial_\mu \alpha)
(A^\mu + \partial^\mu \alpha).
\label{554}
\end{equation}
The Lagrangian remains invariant under the transformations
\begin{equation}
A_{\mu}(x) \rightarrow
A^{\prime}_{\mu}(x) =  A_{\mu}(x) + \partial_{\mu} \Lambda (x), ~~~~~~\alpha (x)  \rightarrow  \alpha '(x) = \alpha (x) -\Lambda (x)
\label{54}
\end{equation}
where $\Lambda (x)$ is an arbitrary scalar function.
The equations of motion for the theory are
\begin{equation}
-\partial^\nu F_{\mu \nu } + m^2 (A_\mu + \partial_\mu \alpha)  = 0,~~~~~ \partial^\mu (A_\mu + \partial_\mu \alpha) = 0.
\label{55}
\end{equation}
One must notice that  by
operating $\partial_\mu $ on the first equation in (\ref{55}) one yields the second. Hence the
latter is consequence of the former. This implies that the gauge transformation
 of the $\alpha $-field can be deduced by knowing that of the $A^\mu$-field.
Similar to (\ref{9}), here we adopt  the ansatz $A_{\mu}(x) = \varepsilon_\mu \exp (ik.x)$ and $\alpha (x)
= \tilde{\alpha}(k) \exp (ik.x)$.  In terms
of the polarization vector $\varepsilon_\mu (k)$, the equations of motion (\ref{55}) become respectively,
\begin{equation}
k^\nu(k_\mu \varepsilon_\nu - k_\nu \varepsilon_\mu ) + m^2(\varepsilon_\mu + ik_\mu \tilde{\alpha}) = 0, ~~~
ik^\nu(\varepsilon_\nu + ik_\nu \tilde{\alpha}) = 0.
\label{566}
\end{equation}
For massless excitations $k^2 = 0$, the second equation in (\ref{566})
gives the Lorentz condition  $ k_\nu \varepsilon^\nu = 0$ which
when substituted in the first gives,
$
\varepsilon_\mu  =- i k_\mu \tilde{\alpha}.
$
Since this is a solution proportional to the 4-momentum $k_\mu$, it can be gauged away by an appropriate
choice of the gauge. Thus massless excitations are gauge artefacts. For $k^2 = M^2 $ (massive excitations),
the equations of motion (\ref{566}) becomes,
\begin{equation}
(m^2 - M^2) \varepsilon^\mu + k^\mu  k_\nu \varepsilon^\nu + im^2  k^\mu \tilde{\alpha}  = 0, ~~~ \tilde{\alpha} = \frac{ik_\nu \varepsilon^\nu}{M^2}.
\label{58}
\end{equation}
Substituting  the second equation of (\ref{58}) in  the first yields,
\begin{equation}
(m^2 - M^2) \varepsilon^\mu + k_\nu \varepsilon^\nu k^\mu (1 - \frac{m^2}{M^2})
=0.
\label{59}
\end{equation}
Now, (\ref{59}) can be satisfied only if $M= m$. Therefore,  the mass of the excitation is given by $m$ itself and  the rest frame momentum 4-vector of the
theory can be written as $ k^\nu = (m, 0, 0, 0)$. Then, in  the rest frame the second
equation in (\ref{566}) gives
$
\varepsilon_0 = -i m \tilde{\alpha}.
$
Therefore, the polarization vector of $A^\mu (x)$ field in (\ref{554}) can be written as
$\varepsilon^{\mu} = (-i m \tilde{\alpha}, \varepsilon^1, \varepsilon^2, \varepsilon^3)^T$   and
 a gauge transformation with the choice $\Lambda (x) = \alpha (x)$  yields its  maximally reduced form,
\begin{equation}
\varepsilon^{\mu} = (0, \varepsilon^1, \varepsilon^2, \varepsilon^3)^T
\label{62}
\end{equation}
where the free components $\varepsilon^1, \varepsilon^2, \varepsilon^3 $
represents the three physical degrees of freedom in the theory.
One must note  that (\ref{62}) is of the same form as that of the $B\wedge F$ theory polarization vector \cite{bc1}.  Therefore, just as in the case of
$B\wedge F$ theory, the action of representation $D(p,q,r)$ (\ref{53-1}) of
$T(3)$ on the polarization vector  (\ref{62}) amounts to a gauge transformation
 in St\"uckelberg extended Proca theory:
\begin{equation}
\varepsilon^{\mu} \rightarrow \varepsilon^{\prime \mu} = {D^{\mu}}_{\nu}(p,q,r)\varepsilon^{\nu} = \varepsilon^{\mu} + \frac{i}{m}(p\varepsilon^1 +
q \varepsilon^2 + r  \varepsilon^3)k^{\mu}.
\label{622}
\end{equation}
The above transformation can be cast in the form of the momentum space
gauge transformation
\begin{equation}
\varepsilon^{\mu}\rightarrow \varepsilon^{\mu} +
ik^\mu \lambda (k)
\label{6221}
\end{equation}
 (where $\Lambda(x)= \lambda (k) e^{ik.x}$) corresponding
to the field $A(x)$, by choosing the field $ \Lambda(x)$ such that
\begin{equation}
\lambda (k)= \frac{(p\varepsilon^1 +
q \varepsilon^2 + r  \varepsilon^3)}{m}.
\label{6222}
\end{equation}

As mentioned before, it is possible to obtain the gauge transformation property of
$\alpha$ field from that of the $A^\mu$ field.
Consider the second relation in (\ref{58}), i.e.,  $\tilde{\alpha} = \frac{ik_\mu \varepsilon^\mu}{m^2} $
and let $\varepsilon^\mu$ undergo the gauge transformation (\ref{6221}) which has the effect
of making a corresponding transformation in $\alpha$ field as
\begin{equation}
\tilde{\alpha} \rightarrow \tilde{\alpha}'  = \frac{ik_\mu (\varepsilon^\mu + ik^\mu \lambda )}{m^2} =\frac{ik_\mu \varepsilon^\mu}{m^2} - \lambda = \tilde{\alpha} - \lambda  .
\label{62233}
\end{equation}
Here $\lambda$ is given by (\ref{6222}) corresponding to the gauge transformation generated by the
translational group $T(3)$ in the $A^\mu(x)$ field. Notice that the above equation (\ref{62233})
correspond to the second equation in (\ref{54}). We have thus obtained the gauge
transformation generated in the $\alpha$ field by $T(3)$ from that in the $A_\mu (x)$-field. It follows therefore that $\alpha$ field can be gauged away
completely by a suitable gauge fixing condition (unitary gauge) and it does not
appear in the physical spectrum of the theory.

Hence it is obvious that the representation $D(p,q,r)$  of $T(3)$ generates
gauge transformation in the massive vector gauge theory governed  by (\ref{554}).

\subsection{Massive symmetric tensor gauge theory}
Consider the massive and gauge noninvariant Einstein-Pauli-Fierz (EPF) theory
in 3+1 dimension as given by the Lagrangian,
\begin{equation}
{\cal L}_L^{EPF} = \frac{1}{2}h_{\mu\nu} \left[ R^{\mu\nu}_L -
\frac{1}{2} \eta^{\mu\nu} R_L\right] - \frac{\mu^2}{2}\left((h_{\mu\nu})^2
 - h^2 \right).
\label{63}
\end{equation}
Just as the Proca theory (section 4.1) can be made gauge invariant by St\"uckelberg
mechanism, the linearized EPF theory also can be made  gauge invariant by
introducing  an additional vector field $A^\mu$ as follows:
\begin{equation}
{\cal L}_L^{EPF} = \frac{1}{2}h_{\mu\nu} \left[ R^{\mu\nu}_L -
\frac{1}{2} \eta^{\mu\nu} R_L\right] - \frac{\mu^2}{2}\left(\left(h_{\mu\nu} +
\partial_\mu A_\nu + \partial_\nu A_\mu          \right)^2 - \left(h +
2 \partial \cdot A\right)^2\right).
\label{64}
\end{equation}
The theory described by (\ref{64}) is invariant under the gauge  transformations
\begin{equation}
h_{\mu\nu} \rightarrow h'_{\mu\nu} = h_{\mu\nu} + \partial_\mu \Lambda_\nu
+ \partial_\nu \Lambda_\mu,~~~
A_\mu (x)\rightarrow A'_\mu (x) = A_\mu (x) - \Lambda_\mu (x)
\label{66}
\end{equation}
The equations of motion for $h_{\mu\nu}$ and   $A_\mu$ are given respectively by following equations.
$$- \Box h^{\mu\nu} + \partial^\mu \partial_\alpha  h^{\alpha \nu} + \partial^\nu\partial_\alpha  h^{\alpha \mu}
 - \partial^\mu \partial^\nu h + \eta^{\mu\nu}(\Box h -\partial_\alpha
\partial_\beta h^{\alpha \beta})$$
\begin{equation}
 - \mu^2 \left[ (h^{\mu\nu} +
\partial^\mu A^\nu + \partial^\nu A^\mu) - \eta^{\mu\nu}(h + 2 \partial \cdot A)\right] = 0
\label{67}
\end{equation} \begin{equation}
\Box A^\mu + \partial_\nu h^{\nu\mu} -  \partial^\mu h -\partial^\mu (\partial \cdot A)  = 0.
\label{68}
\end{equation}
Analogous to the case of massive vector gauge theory discussed before, the equation of motion
(\ref{68}) for $A^\mu$ can be obtained from (\ref{67}) by applying the operator $\partial_\nu$.
Therefore, gauge transformation of $A^\mu$  is obtainable by knowing the gauge transformation of
the $h^{\mu\nu}$ field by a method  similar to the one discussed in section 4.1 for
the case of  St\"uckelberg extended Proca theory.   With   $ h_{\mu \nu}(x) = \chi_{\mu \nu}(k) e^{ik.x}   $  and
$ A_\mu(x) = \varepsilon_\mu (k) e^{ik.x}$
we  now employ the plane wave method to obtain the
maximally reduced polarization
tensor $\chi_{\mu \nu}$   involved in $h_{\mu \nu}$   field. In the momentum space,  the gauge transformations (\ref{66})   can be written as
\begin{equation}
\chi_{\mu\nu} \rightarrow \chi_{\mu\nu}' = \chi_{\mu\nu} + ik_\mu \zeta_\nu
+ ik_\nu \zeta_\mu,~~~~~
\varepsilon_\mu \rightarrow \varepsilon_\mu' = \varepsilon_\mu - \zeta_\mu
\label{70+2}
\end{equation}
(where $\Lambda_\mu (x) = \zeta_\mu (k) \exp (ik.x)$) and the equation of motion  (\ref{67}) for $h_{\mu \nu}$ as
$$k^2\chi^{\mu\nu} - k^\mu k_\alpha \chi^{\alpha \nu} -
k^\nu k_\alpha \chi^{\alpha \mu }
+  k^\mu k^\nu \chi
+\eta^{\mu\nu}(-k^2 \chi +k_\alpha k_\beta \chi^{\alpha \beta})$$
\begin{equation}
-\mu^2 \left[ \chi^{\mu\nu} + ik^\mu\varepsilon^\nu + ik^\nu\varepsilon^\mu
-\eta^{\mu\nu}(\chi + 2ik_\alpha\varepsilon^\alpha) \right] = 0.
\label{71}
\end{equation}
On contracting with $\eta_{\mu\nu}$ and considering only massless ($k^2 = 0$)
excitations,  (\ref{71})reduces to
\begin{equation}
2k_{\mu\nu}\chi^{\mu\nu} + \mu^2\left[3(\chi + 2ik_\mu\varepsilon^\mu)\right] =
0.
\label{72}
\end{equation}
The solution of the above equation is   $\chi^{\mu\nu} = -i(k^\mu\varepsilon^\nu + k^\nu\varepsilon^\mu)$.
Hence it is also the solution of (\ref{71}) with $k^2 = 0$.
It is obvious that this solution is a gauge artefact since one can choose the
arbitrary vector field $\Lambda_\mu = A_\mu$ so as to make this solution vanish.

Next we consider the massive case ($k^2 = M^2, M \neq 0$) and consider the $(00)$
component of the equation of motion (\ref{71})  which, by a straightforward algebra,
can be reduced to
\begin{equation}
{\chi^1}_1 + {\chi^2}_2 + {\chi^3}_3 = 0.
\label{74}
\end{equation}
Similarly the ($0i$) component of (\ref{71}) gives $
{\chi_{0i}} = -iM \varepsilon_i.$
Now, the ($ij$) component of (\ref{71}) is given by
\begin{equation}
k^2\chi_{ij} - \eta_{ij}k^2(\chi - \chi^{00}) - \mu^2[\chi_{ij} - \eta_{ij}
(\chi + 2iM\varepsilon^0)] =0.
\label{76}
\end{equation}
Using (\ref{74}), the above equation can be reduced to
\begin{equation}
k^2\chi_{ij} - \mu^2[\chi_{ij} - \eta_{ij}
(\chi + 2iM\varepsilon^0)] =0
\label{77}
\end{equation}
By adding up the three equations obtained by successively setting
 $i= j = 1, 2, 3$ in (\ref{77}),
 and subsequently using (\ref{74}), we
arrive at
$\chi_{00} = -2 iM\varepsilon_0$.
On the other hand, when $i\neq j$ the equation (\ref{77}) reduces to
\begin{equation}
( \mu^2 - M^2) \chi_{ij} = 0.
\label{79}
\end{equation}
At this juncture notice that only two of the three components $\chi_{ii}, i= 1,
2, 3 $ are
independent on account of the equation (\ref{74}). Also, the $\chi_{00}$ and
$\chi_{0i}$ components can be set equal to zero by choosing the arbitrary
field
$\Lambda_\mu$ to be $A_\mu$. Therefore, if $\chi_{ij} =0$
(for $ i\neq j$) in the above equation (\ref{79}), the number of independent components
of $\chi_{\mu \nu}$ will  be only two.   
Since
this is not the case, we can satisfy the equation (\ref{79})
only if $\mu^2 = M^2 $. Thus we see that the parameter $\mu$  represents the mass of the physical excitations of the field $h_{\mu \nu}$ and that its polarization
tensor is
\begin{equation}
\{ \chi_{\mu \nu} \} = \left( \begin{array} {cccc}
-2 i\mu\varepsilon_0 & -i\mu\varepsilon_1 & -i\mu\varepsilon_2 & -i\mu\varepsilon_3 \\
-i\mu\varepsilon_1   & \chi_{11}        &   \chi_{12}      &  \chi_{13}  \\
-i\mu\varepsilon_2   & \chi_{12}        &   \chi_{22}      &  \chi_{23}  \\
-i\mu\varepsilon_3   & \chi_{13}        &  \chi_{23}       &  \chi_{33} \end{array} \right); ~~~\chi_{11} + \chi_{22} + \chi_{33} = 0.
\label{80}
\end{equation}
As mentioned before, by choosing the  field $\Lambda_\mu$ to be $A_\mu$
and making a gauge transformation, the above form of the polarization tensor can be converted
 to its maximally reduced form given by
\begin{equation}
\{ \chi_{\mu \nu} \} = \left( \begin{array} {cccc}
0 & 0 & 0 & 0 \\
0 &  \chi_{11}        &   \chi_{12}      &  \chi_{13}  \\
0 &  \chi_{12}        &   \chi_{22}      &  \chi_{23}  \\
0 &  \chi_{13}        &  \chi_{23}       &  \chi_{33} \end{array} \right);~~~ \chi_{11} + \chi_{22} + \chi_{33} = 0.
\label{81}
\end{equation}

The action of $D(p, q, r)$ on the polarization tensor
$\{\chi_{\mu\nu}\}$ (\ref{81}) is given by,
$$\{\chi_{\mu\nu}\} \rightarrow  \{\chi_{\mu\nu}\}' =  D(p, q, r) \{\chi_{\mu\nu}\} D^T(p, q, r) = $$
\begin{equation}
\{\chi_{\mu\nu}\} + \left( \begin{array}{cccc}
\left(\begin{array}{c}p(p\chi_{11} + q\chi_{12} + r\chi_{13}) \\
+ q(p\chi_{12} + q\chi_{22} + r\chi_{23}) \\
+ r( p\chi_{13} + q\chi_{23} + r\chi_{33})  \end{array}\right)
&\left(\begin{array}{c} p\chi_{11} + q\chi_{12} \\+ r\chi_{13} \end{array}\right) &
\left(\begin{array}{c}  p\chi_{12} + q\chi_{22} \\ + r\chi_{23} \end{array} \right)&
\left(\begin{array}{c}  p\chi_{13} + q\chi_{23} \\
+ r\chi_{33}  \end{array}\right)  \\
p\chi_{11} + q\chi_{12} + r\chi_{13} & 0 & 0 & 0 \\
p\chi_{12} + q\chi_{22} + r\chi_{23} & 0 & 0 & 0 \\
 p\chi_{13} + q\chi_{23} + r\chi_{33} & 0 & 0 & 0 \\
\end{array} \right)
\label{87+1}
\end{equation}
By choosing
$$ \zeta_0 = \frac{1}{2} (p\zeta_1 + q\zeta_2 + r\zeta_3),~~~
\zeta_1 = \frac{1}{\mu} (p\chi_{11} + q\chi_{12} + r\chi_{13}),~~~
\zeta_2 = \frac{1}{\mu} (p\chi_{12} + q\chi_{22} + r\chi_{23}) $$
$$ \zeta_3 = \frac{1}{\mu}(p\chi_{13} + q\chi_{23} + r \chi_{33});~~~~\chi_{11} + \chi_{22} + \chi_{33} =0 $$
it is straightforward to see that (\ref{87+1}) has the form   of
 the gauge transformation of $\chi_{\mu\nu}$ (see (\ref{70+2})). Notice that when one
makes the choices for  the components $\zeta_1, \zeta_2, \zeta_3$ in terms of
the three parameters $p,q,r$ of the translational group $T(3)$, the component
$\zeta_0$ gets automatically fixed. Therefore, in  the gauge transformation
(\ref{87+1}) generated by the representation $D(p,q,r)$ of $T(3)$ only the
three space components of the arbitrary field $\zeta_\mu$ remain independent where as
for the generation of the  complete set of gauge transformations (\ref{70+2}) all the
four components of $\zeta_\mu$ should be  independent of one another.
Hence, the above gauge transformations (\ref{87+1})  generated by the
translational group $D(p,q,r)$ does not exhaust the complete set of gauge
transformations
available to the massive symmetric tensor gauge theory.   As mentioned before,
the gauge transformation of the $A_\mu$-field can be obtained from that of the $h_{\mu\nu}$-field, though
the former one does not appear in the physical spectrum of the theory.

\subsection{Massive antisymmetric tensor gauge theory}

Here we discuss  the  role  the translational group $T(3)$ as  gauge generator in the St\"uckelberg extended
massive KR theory which is another example of a reducible gauge theory. Though the
analysis in this case closely resembles that of St\"uckelberg extended EPF theory,
here the reducibility  of the gauge transformation  is manifested in the  gauge
generation by $T(3)$.
The Lagrangian of the St\"uckelberg extended
massive KR theory is 
\begin{equation}
{\cal L} = \frac{1}{12}H_{\mu\nu\lambda}H^{\mu\nu\lambda} - \frac{m^2}{4} (B_{\mu\nu} + \partial_\mu A_\nu -
\partial_\nu A_\mu ) (B^{\mu\nu} + \partial^\mu A^\nu -
\partial^\nu A^\mu ).
\label{n1}
\end{equation}
 It can be easily verified that this theory
is invariant under the joint gauge transformations
\begin{equation}
B_{\mu\nu}(x) \rightarrow B_{\mu\nu}(x) + \partial_\mu F_\nu
(x) - \partial_\nu F_\mu (x),~~~~
A_\mu (x)\rightarrow A_\mu (x) - \Lambda_\mu (x).
\label{n3}
\end{equation}
Here we must notice that gauge transformation of $B_{\mu\nu}$ is  reducible exactly
as  that in massless KR theory.
The equations of motion (corresponding to $B_{\mu\nu}$  and $A_\mu$) following from (\ref{n1}) are given by
\begin{equation}
\partial_\mu H^{\mu\nu\lambda} + m^2 (B^{\nu\lambda} + \partial^\nu A^\lambda
- \partial^\lambda  A^\nu ) = 0 
\label{n4}
\end{equation}
\begin{equation}
\partial_\mu (B^{\mu\nu} + \partial^\mu A^\nu - \partial^\nu A^\mu ) = 0.
\label{n5}
\end{equation}
As in the case of the St\"uckelberg extended massive theories considered 
previously in sections 4.1 and 4.2, the equation of motion (\ref{n5}) for
$A^\nu$ can be obtained from the equation (\ref{n4})  by the 
 application of the operator $\partial_\lambda$. Hence,  one can  easily obtain
the gauge
transformation  of the $A^\nu$ field from that of the
$ B^{\nu\lambda}$ field.

In order to obtain the maximally reduced polarization tensor
$\varepsilon^{\mu\nu}(k)$ corresponding to the antisymmetric field
$B^{\mu\nu}(x)$, as usual we use the ansatz $
B^{\mu\nu}(x) = \varepsilon^{\mu\nu}(k) e^{ik\cdot x},~~A^{\mu}(x) = \varepsilon^{\mu}(k)  e^{ik\cdot x}$
and employ the plane wave method.
The momentum space gauge transformation of  $\varepsilon^{\mu\nu}(k)$
now has   the same form as (\ref{18+1}).
The equation of motion
 (\ref{n4}) can be written (in the momentum space) as
\begin{equation}
-k^2 \varepsilon^{\nu\lambda} - k^\nu k_\mu \varepsilon^{\lambda\mu} - k^\lambda k_\mu \varepsilon^{\mu \nu} +
m^2 (\varepsilon^{\nu\lambda} + ik^\nu \varepsilon^{\lambda} - ik^\lambda \varepsilon^{\nu}) = 0.
\label{n7}
\end{equation}
If $k^2 = 0$ (massless excitations), the above equations reduces to
\begin{equation}
- k^\nu k_\mu \varepsilon^{\lambda\mu} - k^\lambda k_\mu \varepsilon^{\mu \nu} + m^2 (\varepsilon^{\nu\lambda} +
ik^\nu \varepsilon^{\lambda} - ik^\lambda \varepsilon^{\nu}) = 0
\label{n8}
\end{equation}
the most general solution for which is
\begin{equation}
\varepsilon^{ \nu\lambda}(k) = C(ik^\nu \varepsilon^{\lambda} - ik^\lambda
\varepsilon^{\nu}) + D (\epsilon^{\nu\lambda\tau\sigma}
k_\tau\varepsilon_\sigma )
\label{n9}
\end{equation}
where $C$ and $D$ are constants to be fixed.
Substituting this solution (\ref{n9}) in (\ref{n8}), we can easily see that
$C = -1$ and $D = 0$.
Therefore, the solution to (\ref{n7}) corresponding to  massless excitations
is $
\varepsilon^{ \nu\lambda}(k) = -ik^\nu \varepsilon^{\lambda} + ik^\lambda
\varepsilon^{\nu}.$
However, such solutions can be gauged away by choosing the arbitrary field $\Lambda^\mu (x)  = A^\mu (x)$ which shows
that massless excitations are gauge artefacts.

Next we consider the massive case, $k^2 = M^2,~(M\neq 0)$ where it is
possible to go to the rest frame where
$k^\mu = (M, 0,0,0)^T$.  In the rest frame,
the equation of motion (\ref{n7})
reduces to
\begin{equation}
(m^2 -M^2) \varepsilon^{ \nu\lambda} -M(k^\nu \varepsilon^{\lambda 0} + k^\lambda \varepsilon^{ 0\nu }) + m^2 (ik^\nu \varepsilon^{\lambda} - ik^\lambda
\varepsilon^{\nu}) = 0.
\label{n11}
\end{equation}
Note that, since the polarization tensor $\varepsilon^{ \nu\lambda}$ is antisymmetric,
all its diagonal entries are automatically zero. Considering the components
of (\ref{n11}) for which $(\nu =0, \lambda = i)$, we have   $
 \varepsilon^{ i0} = iM \varepsilon^{ i}.$
For $(\nu =i, \lambda = j)$ with $i\neq j$, the equation (\ref{n11}) gives
$m^2 -M^2) \varepsilon^{ij} = 0. $
This leads to two possibilities; either $\varepsilon^{ij} = 0$ or $ M^2 = m^2$.
The former possibility can be ruled out by the following reasoning.
Since (\ref{n1}) is the first class version (obtained by a St\"uckelberg
extension mechanism)
of massive KR theory possessing three physical degrees of
freedom \cite{bb}, the former too must inherit the same number of degrees of freedom.
However, the $\varepsilon^{ i0}$ elements can all be made to vanish by the gauge choice $\Lambda_\mu = A_\mu$. Therefore the possibility
 $\varepsilon^{ij} = 0$ leads to a null theory and hence should be discounted.
Therefore, we have $M^2 = m^2$ which  is also consistent with the degrees of
freedom counting. Finally, analogous to (\ref{81}), the maximally reduced
form of the polarization tensor corresponding to (\ref{n1}) is given by,
\begin{equation}
\{\varepsilon^{\mu\nu}\} = \left( \begin{array} {cccc}
0 & 0 & 0 & 0 \\
0 &  0        &   \varepsilon^{12}      &  \varepsilon^{13}  \\
0 &     -\varepsilon^{12}     &     0    &  \varepsilon^{23}  \\
0 &   -\varepsilon^{13}      &  -\varepsilon^{23}       &  0 \end{array} \right)
\label{n14}
\end{equation}
As in the case of St\"uckelberg extended EPF theory, the $A_\mu$-field  disappears from the physical spectrum in this case also.
Here it must be emphasized that the maximally reduced polarization tensor
of $B\wedge F$ theory also has the same form (\ref{n14}). This is not
surprising since the physical sector of $B\wedge F$ theory is equivalent to
massive KR theory whose first class version is the theory (\ref{n1}) under
consideration now \cite{bb}.
We now study the action of
$D(p, q, r)$ on (\ref{n14}) is given by,
$$\{\varepsilon^{\mu\nu}\}\rightarrow \{\varepsilon^{\mu\nu}\}' = D(p, q, r) \{\varepsilon^{\mu\nu}\} D^T(p, q, r)$$
\begin{equation}
=\{\varepsilon^{\mu\nu}\} + \left( \begin{array} {cccc}
0 & -q\varepsilon^{12} - r \varepsilon^{13} & p\varepsilon^{12} - r \varepsilon^{23} & p\varepsilon^{13} + q \varepsilon^{23} \\
q\varepsilon^{12} + r \varepsilon^{13} & 0 & 0& 0 \\
-p\varepsilon^{12} + r \varepsilon^{23}& 0 & 0& 0 \\
- p\varepsilon^{13} - q \varepsilon^{23}& 0 & 0& 0 \\
\end{array} \right)
\label{n15}
\end{equation}
This can be considered to be the gauge 
transformations of $\varepsilon^{ \nu\lambda}$ (\ref{18+1}) if we   choose
\begin{equation}
f^1 =\frac{1}{m}(q\varepsilon^{12} + r \varepsilon^{13}),~~~
f^2 =  \frac{1}{m}(-p\varepsilon^{12} + r \varepsilon^{23}), ~~~
f^3 = \frac{-1}{m}( p\varepsilon^{13} + q \varepsilon^{23})
\label{n16}
\end{equation}
Note that the component $f^0$ remains completely
undetermined and does not depend at all either on the parameters $p,q,r$ of
$T(3)$ or on the maximally reduced polarization tensor of the theory
whereas the other components $f^1, f^2, f^3$ are determined by these
parameters and the elements of the polarization tensor.  Hence, $T(3)$ generates the complete set of gauge transformations in
the St\"uckelberg extended massive KR theory.
Interestingly, it is exactly in the same fashion as in the present case (of 
St\"uckelberg  extended massive KR theory) that gauge transformations 
of $B\wedge F$ theory are generated by the translational group $D(p,q,r)$ (we
refer to \cite{bc1} for details).  Analogous to the gauge transformation
generated by $W(p,q)$ in massless KR theory, for any given set of ($f^1, f^2, f^3$)
we have a continuum of values for $f^0$, representing the reducibility of
the gauge transformation in the underlying 2-form field  both in  St\"uckelberg
extended first class version of massive KR theory and in the $B\wedge F$ theory.
Therefore, the complete independence of the time-component of $f^\mu$ on the 
maximally reduced polarization tensor or on the parameters of the group 
$D(p,q,r)$ is a consequence of the reducibility of the gauge transformations 
of these theories.

Analogous to the hierarchical structure involving the elements of $T(2)$ 
present in the gauge generation in massless $n$-form theories, there is 
a hierarchical structure in the gauge transformations generated by $T(3)$ 
in massive $n$-form theories also. In section 4.1 we have seen that an element
of $T(3)$ generates gauge transformation in massive 1-form theory (the 
St\"uckelberg extended Proca theory). In the massive 2-form  (the
St\"uckelberg extended massive KR) theory, two elements of $T(3)$ are involved
as generators of gauge transformation, one element for the gauge transformation
of the field $B^{\mu\nu}$ and a second element for the `gauge transformation'
$F^\mu \rightarrow F^\mu + \partial^\mu \beta$ which correspond to the 
reducibility of gauge freedom  in $B^{\mu\nu}$. (This transformation is of the same form as
the first transformation in (\ref{54}) corresponding to massive vector theory
and hence may be considered to be generated by $T(3)$.)

\section{Dimensional Descent}

Dimensional descent \cite{bc2} is a method by which one can obtain the
 energy-momentum
vector, polarization tensor and the gauge generating representation of the translational
subgroup of Wigner's little group etc in a  massive gauge theory living in a
certain space time
dimension from similar results for gauge theories in one higher dimension. In this sense,
dimensional descent is a unification scheme for the results presented in the previous sections.
A closely related concept is the idea of `dimensional reduction' by which massless theories in a
given  dimension can be related to massive theories in a space of one lower dimension as can be seen from
\cite{ms}.
Similar ideas are  used in the context
of string theory also where a massive particle is viewed as a massless particle in one higher dimension
with the mass being considered as the momentum component along the additional dimension \cite{blt}.

We begin our discussion of dimensional descent by noting that, the translational group $T(3)$ which generates gauge transformation
in 3+1 dimensional  $B\wedge F$ theory and in the massive excitations of St\"uckelberg extended Proca and EPF theories, is an invariant
 subgroup of $E(3)$.  Now, just as $E(2)$ is the generator of gauge transformation in 4-dimensional
Maxwell theory, $E(3)$ generates gauge transformation in 5-dimensional Maxwell theory. This
indicates that the generators of gauge transformations in the above mentioned 
massive gauge theories and  5-dimensional Maxwell theory 
are related.

An element of Wigner's little group in 5 dimensions  \cite{bc2}
can be written as
\begin{equation}
W_5(p,q,r; \psi, \phi, \eta) =
\left( \begin{array}{ccccc}
1+ \frac{p^2 + q^2 + r^2}{2} & p & q & r & -\frac{p^2 + q^2 + r^2}{2} \\
p & & & & -p \\
q & &R(\psi, \phi, \eta) & & -q \\
r & & & & -r  \\
\frac{p^2 + q^2 + r^2}{2} &  p & q & r & 1 -\frac{p^2 + q^2 + r^2}{2}
\end{array}\right)
\label{88}
\end{equation}
where $p,q,r$ are  any real numbers, while $R(\psi, \phi, \eta) \in
SO(3)$,  with $(\psi, \phi, \eta)$ being a triplet of Euler angles. The
corresponding element of the translational group $T(3)$ can be trivially
obtained by setting $R(\psi, \phi, \eta)$ to be the identity matrix and
will be denoted by $W(p,q,r) = W_5(p,q,r; 0)$. \\
 
Let us now consider free Maxwell theory in 5-dimensions ($
{\cal L} = -{\frac{1}{4}}{F^{ij}F_{ij}},~ i
,j = 0, 1, 2, 3, 4$)
For a photon of energy $\omega$ (in 5-dimensional space-time) propagating in the $i = 4$ direction,
the momentum 5-vector is given by
\begin{equation}
k^i = (\omega, 0, 0, 0, \omega)^T.
\label{90}
\end{equation}
By following the plane wave method and proceeding exactly as in section 2, one can show that the
maximally reduced form of the polarization vector of the photon is
\begin{equation}
\varepsilon^{i} = (0, \varepsilon^1,\varepsilon^2,\varepsilon^3 , 0)^T
\label{91}
\end{equation}
where $\varepsilon^1,\varepsilon^2,\varepsilon^3$ represent the three 
transverse degrees of freedom
(since the polarization vector satisfies the `Lorentz gauge'
$\varepsilon^{i}k_i = 0$).
If we now suppress the last row of the column matrices  $k^{i}$ (\ref{90}) and $\varepsilon^{i}$ (\ref{91}), we end up respectively with the energy-momentum 4-vector and the polarization vector 
 of St\"uckelberg extended Proca model in 3+1 dimensions. This is equivalent to applying
the projection operator given by the matrix
\begin{equation}
{\cal P}= diag(1, 1, 1, 1, 0)
\label{92}
\end{equation}
to the momentum 5-vector (\ref{90}) and the polarization vector (\ref{91}).
Similarly, it is  possible to derive the polarization tensor of
St\"uckelberg extended EPF theory (\ref{64}) from that  of linearized  Einstein gravity
 in 5-dimensions by a  procedure analogous to the one
  described above. As done in the 3+1 dimensional case, one can easily 
show that the maximally reduced form of the polarization tensor of 4+1
dimensional linearized gravity is 
\begin{equation}
\{ \chi^{ij} \} = \left( \begin{array} {ccccc}
0 & 0 & 0 & 0 & 0 \\
0 &  \chi^{11}        &   \chi^{12}      &  \chi^{13}  & 0 \\
0 &  \chi^{12}        &   \chi^{22}      &  \chi^{23}   & 0 \\
0 &  \chi^{13}        &  \chi^{23}       &  \chi^{33}   & 0 \\
0 & 0 & 0 & 0 & 0 \end{array} \right); ~~~~ \chi^{11} + \chi^{22} + \chi^{33} =0.
\label{94}
\end{equation}
By suppressing the
last row and  column of the polarization tensor(\ref{94}),
one obtains the polarization tensor (\ref{81})  of the  St\"uckelberg extended
EPF model in  3+1
dimensions. For this purpose, consider the  action of $W(p, q, r)$ on
$\varepsilon^{i}$.
\begin{equation}
\varepsilon^{i} \rightarrow {\varepsilon'^{i}} = \varepsilon^{i} + \delta \varepsilon^{i} =
{{W_5 (p, q, r)}^{i}}_{j} \varepsilon^{j} = \varepsilon^{i} +
( p\varepsilon^1 + q\varepsilon^2 + r\varepsilon^3)\frac{k^i}{\omega}
\label{95}
\end{equation}
This is indeed a gauge transformation in (4+1) dimensional Maxwell theory.
Applying the projection operator ${\cal P}$  (\ref{92}) on (\ref{95}) yields
\begin{equation}
\delta {\varepsilon}^{\mu} = {\cal P}\delta\varepsilon^{i}
= \frac{1}{\omega}( p\varepsilon^1 + q\varepsilon^2 + r\varepsilon^3)k^{\mu}
\label{96}
\end{equation}
Here ${\varepsilon}^{\mu} = (0,\varepsilon^1,\varepsilon^2,\varepsilon^3 )^T$ corresponds to the
polarization vector in the St\"uckelberg extended Proca theory and $k^\mu$ is the
 momentum vector of a particle at rest in 3+1 dimensions. (Here, the time component $\omega$
of a 5-dimensional a massless particle, moving along the extra 5th dimension
is identified with the mass $\omega$ of a massive particle at rest
in 4-dimensional space-time.)
Modulo an $i$ factor, this is precisely how the polarization vectors  in the
massive gauge theory (\ref{554}) transforms under gauge transformation (see Section 4.1).
The form of (\ref{96}), makes it obvious that 
\begin{equation}
\delta \varepsilon^{\mu} = D(p,q,r) \varepsilon^{\mu} - {\varepsilon}^\mu
\label{97}
\end{equation}
where
$D(p, q, r)$ is given by (\ref{53-1}). Thus, in this fashion we are able to 
derive  the gauge generating representation $D(p, q, r)$ of $T(3)$ in a massive vector field by a
judicious application of
the projection operator ${\cal P}$  (\ref{92}) from the gauge transformation
relation of a higher dimensional massless gauge theory.

We now consider  the polarization matrix $\{\chi^{ij}\} $ (\ref{94}) of 
linearized gravity in 5-dimensions\footnote{In this regard, we recollect
a comment made in \cite{sc} that translational subgroup of Wigner's little group
for massless particles generates gauge transformations only in the 3+1
dimensional version of linearized gravity, but not in its higher dimensional
versions. This was mistakenly ascribed to the mismatch in the number of
degrees of freedom ($\frac{d(d-3)}{2}$) in higher dimensional linearized gravity and the
number of parameters ($d-2$) of the translational subgroup of Wigner's little group for massless particles in $d> 4$.
However, this need to be amended as this mismatch is
of no consequence in this regard and it must be stated that the translational
subgroup generates gauge transformations for linearized gravity any dimension $d\geq 4$.}
for which  the transformation under the action of $W(p,q,r)$  is given by,
$$\{\chi^{ij}\} \rightarrow \{\chi '^{ij}\} = W(p, q, r)\{\chi^{ij}\} W^T (p, q, r)
=\{\chi^{ij}\} + \delta \{\chi^{ij}\}$$
where,
\begin{equation}
\delta \{\chi^{ij}\} =  \{\delta \chi^{ij} \}
= \left(
\begin{array}{ccccc}
ap+ bq+rc & a & b & c & ap+ bq+rc \\
a & 0 & 0 & 0 & a \\
b & 0 & 0 & 0 &  b \\
c & 0 & 0 & 0 & c\\
ap+ bq+rc & a & b & c & ap+ bq+rc
\end{array}
\right)
\label{98}
\end{equation}where,
$a = p\chi^{11}+q\chi^{12}+ r\chi^{13},~ b = p\chi^{12}+q\chi^{22}+
r\varepsilon^{23}$ and
$c = p\chi^{13}+q\chi^{23}+ r\chi^{33}$ with $ \chi^{11} + \chi^{22} +\chi^{33}=0$.
Again this can be easily recognized as a gauge transformation\footnote{Exactly
 as in the case of 4-dimensional case, this gauge transformation forms only a
subset of the full set of gauge transformation available in 5-dimensional linearized gravity.}  in (4+1)
dimensional linearized gravity involving massless quanta, as $\delta \chi^{ij}$ can be expressed as $ (k^{i} \zeta^{j}(k) + k^{j}\zeta^{i}(k))$
with suitable choice for $\zeta^{i}(k)$, where $k^i = (\omega , 0, 0, 0, \omega )^T$.
By applying the projection operator
${\cal P}$ on (\ref{98}), we get
the change (under gauge transformation) in the 3+1 dimensional polarization matrix
$ \{ {\chi}^{\mu \nu}\}$ (of St\"uckelberg extended EPF theory), by the formula,
$\delta \{ {\chi}^{\mu \nu}\} = {\cal P}\delta \{\chi^{ij}\}{\cal P}^T$. This simply
amounts to a deletion of the last row and column of $\delta \{\chi^{ij}\} $.
The result
can be expressed more compactly as
\begin{equation}
\delta \{\chi^{\mu\nu}\} = (D\{\chi^{\mu \nu}\}D^T - \{\chi^{\mu \nu}\})
\label{99}
\end{equation}
where $D=D(p,q,r)$  (see (\ref{53-1})).
This has the precise form of gauge transformation of the
 polarization matrix
of  St\"uckelberg extended EPF model, since it can be cast in the form
\begin{equation}
\delta {\chi}^{\mu \nu} = (k^{\mu} \zeta^{\nu}(k) + k^{\nu}\zeta^{\mu}(k))
\label{100}
\end{equation}
for a suitable $\zeta^{\mu}(k)$, where $k^{\mu} = (\mu, 0, 0, 0)^T$. Here we have identified $\omega$ with $\mu$.
 
Clearly the
 generators $T_1 = \frac{\partial D}{\partial p}, T_2 =
\frac{\partial D}{\partial q}, T_3 = \frac{\partial D}{\partial r} $ provide a commuting
Lie algebra basis for the group $T(3)$.
One can easily verify
\begin{equation}
 D(p, q, r) = e^{pT_1 + qT_2 + rT_3} = 1 + pT_1 + qT_2 + rT_3
\label{103}
\end{equation}
so that the change in the polarization vector  $\varepsilon^\mu$ can be expressed
as the action of a Lie algebra element,
\begin{equation}
\delta \varepsilon^\mu = (pT_1 + qT_2 + rT_3) \varepsilon^\mu
\label{103+1}
\end{equation}
Besides this, $D(p,q,r)$ also preserves the 4-momentum of a particle at rest.
Thus we have shown, how this representation
(\ref{53-1}) of $T(3)$ can be connected to the Wigner's little group for
massless particle in 4+1 dimension through appropriate projection in the
intermediate steps, where the massless particles moving in  4+1 dimensions
can be associated with a massive particle at rest in 3+1 dimensions.

The method of dimensional descent as applied in the case of 3+1 dimensional
$B\wedge F$ theory
was earlier discussed in \cite{bc2} where it was shown that one can arrive at
the representation $D(p,q,r)$ of translational group $T(3)$ by considering
the gauge transformation properties of 4+1 dimensional massless KR theory.
Since the physical sectors of St\"uckelberg extended massive KR theory
and $B\wedge F$ theory are equivalent (and hence possess identical
  rest frame momentum 4-vectors and maximally reduced
polarization tensors), it is possible to obtain the gauge generating
representation $D(p,q,r)$ of $T(3)$ for St\"uckelberg extended
massive KR theory using dimensional descent proceeding exactly as
was done in \cite{bc2} for the case of $B\wedge F$ theory.

\section{Conclusion}
 The results of study can be summarized as follows.
We have shown that the
representation of the translational  $T(3)$ that acts as a gauge generator in
the topologically massive  $B\wedge F$  gauge theory also generate gauge transformations in the
St\"uckelberg extended gauge invariant versions of 3+1 dimensional Proca, Einstein-Pauli-Fierz and massive Kalb-Ramond theories. This representation of
$T(3)$ along with the polarization vectors/tensors and the momentum 4-vectors of these St\"uckelberg extended theories are derived systematically
using the method of dimensional descent by starting from the appropriate massless gauge theories living in 4+1 dimensional space-time.
Also, we have  reexamined  the gauge generations in 3+1 dimensional massless tensor gauge theories  (linearized gravity and Kalb-Ramond theories)
by  the  translational group $T(2)$ and showed that
gauge transformations generated by $T(2)$ in these theories forms only a subset of the whole  spectrum
 of gauge transformations available.
Similarly the gauge generation by $T(3)$ in the St\"uckelberg
extended Einstein-Pauli-Fierz theory is   also partial. However, in $B\wedge F$ and  St\"uckelberg extended  massive Kalb-Ramond  theories,
 full set of gauge transformations are generated by $T(3)$. In this connection, we have clarified several subtle points
concerning the gauge generation by translational groups.
 It should be emphasized that  in the  case of reducible gauge systems (massless Kalb-Ramond theory,
$B\wedge F$  theory and the St\"uckelberg extended massive Kalb-Ramond theory) the gauge generation by the relevant translational groups
manifestly exhibits the reducibility of the gauge transformations. Further more,
 a hierarchical structure is noticed in the gauge generation by translational groups in both massive and massless theories having
 $n$-form fields as their basic gauge fields, namely $n$ independent elements of the respective  translational group
being involved in the gauge generation in a  $n$-form gauge theory.

{\bf Acknowledgment:} Author is thankful to Dr. R. Banerjee for  suggesting
this  problem and for  several useful comments. 
Thanks are also due to Dr. B. Chakraborty for many
 illuminating discussions regarding this work. A part of the work is done with the support of a Visiting Fellowship from the Saha Institute of
 Nuclear Physics.

\end{document}